\documentclass[9pt, sigconf]{acmart}
\usepackage[compact]{titlesec} 
\titlespacing{\section}{1pt}{1pt}{1pt} 
\titlespacing{\subsection}{1pt}{1pt}{1pt} 
\titlespacing{\subsubsection}{1pt}{1pt}{1pt} 
\usepackage{booktabs}
\usepackage{balance}
\usepackage{multirow}
\usepackage{xspace}
\usepackage{fancyhdr}
\usepackage{graphicx}
\usepackage{epstopdf}
\usepackage{tabularx}
\usepackage{adjustbox}
\usepackage[utf8x]{inputenc} 
\usepackage{url}
\usepackage[ruled,vlined,linesnumbered,noresetcount,titlenumbered]{algorithm2e}
\usepackage{color}
\definecolor{Orange}{rgb}{0.9,0.5,0}
\definecolor{NavyBlue}{rgb}{0.1, 0.4, 0.8}
\definecolor{Magenta}{rgb}{0.8, 0.1, 0.6}
\definecolor{Green}{rgb}{0.1, 0.8, 0.3}
\definecolor{DarkGreen}{rgb}{0.0, 0.7, 0.2}
\definecolor{Brown}{rgb}{0.4, 0.3, 0.1}
\definecolor{Burgundy}{rgb}{0.5, 0.0, 0.13}
\definecolor{BrightCerulean}{rgb}{0.11, 0.67, 0.84}
\definecolor{BlueViolet}{rgb}{0.33,0.1,0.5}

\begin{document}

\renewcommand\footnotetextcopyrightpermission[1]{}
\pagestyle{plain}
\copyrightyear{} 
\acmYear{} 
\acmConference[]{}{}{}
\acmBooktitle{}
\acmPrice{}
\acmDOI{}
\acmISBN{}
\settopmatter{printacmref=false}
\renewcommand{\abstractname}{ABSTRACT}
\renewcommand{\refname}{REFERENCES}
\title{Privacy-preserving Voice Analysis via \\Disentangled Representations} 

\author{Ranya Aloufi, Hamed Haddadi, David Boyle}
\affiliation{%
  \institution{Systems and Algorithms Laboratory}
  \institution{Imperial College London}}

\begin{abstract}
Voice User Interfaces (VUIs) are increasingly popular and built into smartphones, home assistants, and Internet of Things (IoT) devices. Despite offering an always-on convenient user experience, VUIs raise new security and privacy concerns for their users. In this paper, we focus on attribute inference attacks in the speech domain, demonstrating the potential for an attacker to accurately infer a target user's sensitive and private attributes (e.g. their emotion, sex, or health status) from deep acoustic models. To defend against this class of attacks, we design, implement, and evaluate a user-configurable, privacy-aware framework for optimizing speech-related data sharing mechanisms. Our objective is to enable primary tasks such as speech recognition and user identification, while removing sensitive attributes in the raw speech data before sharing it with a cloud service provider. We leverage \emph{disentangled representation learning} to explicitly learn independent factors in the raw data. Based on a user's preferences, a supervision signal informs the filtering out of invariant factors while retaining the factors reflected in the selected preference. Our experimental evaluation over five datasets shows that the proposed framework can effectively defend against attribute inference attacks by reducing their success rates to approximately that of guessing at random, while maintaining accuracy in excess of 99\% for the tasks of interest. We conclude that negotiable privacy settings enabled by disentangled representations can bring new opportunities for privacy-preserving applications.
\end{abstract}

\ccsdesc[500]{Embedded systems}
\ccsdesc[300]{Voice-enabled}
\ccsdesc{Security and Privacy}
\ccsdesc[100]{Performance and Utility}

\keywords{Speech Analysis, Voice Synthesis, Voice Privacy, Internet of Things (IoT)}
\maketitle

\section{\uppercase{Introduction}}
\label{sec:introduction}

\begin{figure*}[t!]
  \centering
  \includegraphics[width=\textwidth,height=8.5cm]{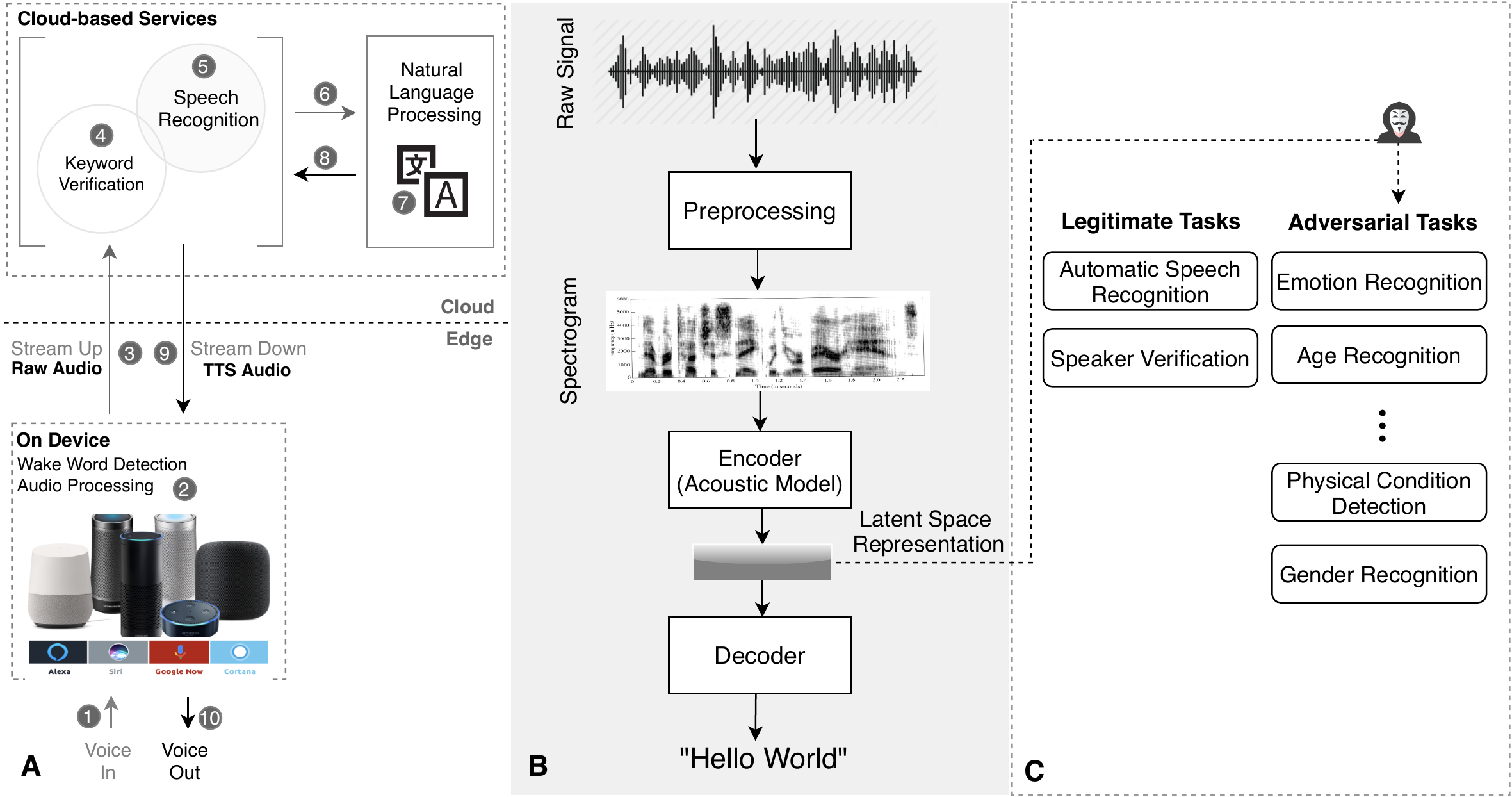}
  \caption{(A) Voice-controlled Systems, (B) End-to-End Automatic Speech Recognition Systems, (C) Potential Attribute Inference Attacks}
  \Description{}
        \label{fig:framework}
\end{figure*}

Voice-controlled IoT devices and smart home assistants have gained huge popularity on our devices and in our households. Intuitive interaction between users and services is enabled by analyzing speech signals. For example, smart assistants (e.g., Google Assistant, Amazon Echo, and Apple Siri) and voice browsing (e.g., Google Search) use Voice User Interfaces (VUIs) to activate the voice assistant to control IoT devices or perform tasks such as browsing the Internet and/or making recommendations. Figure~\ref{fig:framework} (A) shows an overview of how these systems work. Although devices often suffer from frequent false activations~\cite{PETS_2020smarthome}, it all begins with some kind of trigger such as `Okay, Google', `Alexa', and `Hey, Siri' to inform the system that speech-based data will be received. Once a voice stream is captured by a device, it outsources analysis to cloud services such as automatic speech recognition (ASR), speaker verification (SV), and natural language processing (NLP) where higher performance is achievable. This frequently involves communicating instructions to other connected devices, appliances, and third-party systems. Finally, text-to-speech services are often employed in order to speak back to the user. Our voice signal is a rich source of personal and sensitive data. It contains indicators of a variety of emotions, physical and mental health and well-being, etc., and thus raises unprecedented security and privacy concerns where raw data or models derived thereof are transmitted to third parties. The signal contains linguistic and paralinguistic information such as age, gender, health status, personality, friendliness, mood, and emotions~\cite{schuller1988emotion}.

Today, deep learning models are playing a pivotal role in speech signal processing to enable natural and intuitive communication with our smart devices. For example, recent end-to-end (E2E) automatic speech recognition systems rely on autoencoder architecture as a way of folding separate acoustic models, pronunciation, and language models (AM, PM, LM) of a traditional ASR system into a single neural network~\cite{chan2016listen, chiu2018state, wang2019end, hannun2014deep}, as shown in Figure~\ref{fig:framework} (B). These models train by ingesting speech spectrograms as alternative frequency-based representations for speech signals and generate text transcriptions. The encoder encodes the input acoustic feature sequence into a vector, which encapsulates the information for its input to help the decoder in predicting the sequence of symbols. Although these models have comparable performance with conventional models~\cite{chiu2018state}, they have been designed without considering potential privacy vulnerabilities, given the need to train on real voice data, which contains a significant amount of sensitive information. 

Attribute inference attacks may aim to reveal individuals' sensitive attributes (e.g. emotion, gender, health status, etc.) that they did not intend or expect to share. Several privacy violations may arise by obtaining these sensitive data without individuals' awareness or permission. In this paper, we focus on an adversarial privacy leakage scenario of deep representations for speech processing tasks. In particular, we focus on the probability of inferring sensitive attributes using deep acoustic models that perform different operations like speech to text translation or speaker recognition. For example, an attacker may use an acoustic model trained for speech recognition or speaker verification to learn further sensitive attributes from user input even if not present in its training data, as shown in Figure~\ref{fig:framework} (C). The attacker may use the output of these models to train classifiers to infer private attributes. We can measure an attack's success as the increase in inference accuracy over random guessing~\cite{yeom2018privacy,wagner2018technical}, and we find that an attacker can achieve high accuracy in inferring sensitive attributes, ranging from 40\% to 99.4\%, which is three or four times better than guessing at random, depending on the acoustic conditions of the input. We discuss this further in Section~\ref{sec:evaluation.AE}.

In order to limit the success of such attacks, we propose a user-driven framework designed to offer a practical defense against attribute inference attacks. A challenge in designing the proposed framework is to consider individuals' privacy preferences (i.e., giving users a choice with regards to their privacy~\cite{kolter2010user}) in sharing their data. More precisely, different users may have varying privacy preferences as to the type of analytics that can be done on their data, perhaps depending on the devices and services with which they are interacting. For instance, when contacting a health service provider, a user may prefer to share raw data without altering it, whereas a user may prefer to filter sensitive data when potentially interacting with advertising companies. To address this challenge, the proposed framework works in two phases. In Phase I, the user selects their privacy preference, where each of the preferences is associated with a set of tasks (e.g. speech recognition) that can be performed on the their data. In Phase II, we take advantage of \emph{learning disentangled representation}~\cite{van2017neural} in the observed data to explicitly derive each dimension to reflect independent factors for a particular task.

Finally, we evaluate the proposed framework's efficiency against this class of attacks using various datasets, which were recorded under different acoustic conditions (IEMOCAP~\cite{busso2008iemocap}, RAVDESS~\cite{RAVDESS}, SAVEE~\cite{haq2008audio}, LibriSpeech~\cite{panayotov2015librispeech}, and VoxCeleb~\cite{nagrani2017voxceleb}) to simulate the real-time environment in which voice recordings are collected. The results show the effectiveness of our proposed framework in reducing the success rate of the attacker to less than or equal to randomly guessing for identifying sensitive attributes.

\textbf{Contribution.} Our contributions can be summarized as follows:
\begin{itemize}
\item We show the vulnerability of underlying acoustic models used by speech processing tasks under attribute inference attack scenarios. Models' predictions may exploit such models to learn further information about users. We measure the success of these attacks by the increase in inference accuracy over random guessing. We demonstrate the importance of developing privacy-preserving solutions that can run at the edge, i.e. before sharing data with service providers.
\item We propose and develop a privacy-aware, configurable defence framework against attribute inference attacks. We design it to include users' privacy preferences in managing the privacy-utility trade-off inherent in data sharing. Precisely, we allow a user to explicitly adjust the disentangled representation of his/her preference, learned by the framework from his/her data. According to our experimental results, we conclude that the controllability enabled by the disentanglement may define a new direction in developing privacy-preserving applications that satisfy the transparency principle.
\item We experimentally evaluate the proposed framework over various datasets, and the results show its effectiveness in confronting this type of attack by filtering the sensitive attributes while maintaining high accuracy, i.e. >99\%, for the tasks of interest. Audio snippets are available online\footnote{YouTube:~\href{https://tinyurl.com/y932f37m}{https://tinyurl.com/y932f37m}} and our code is open source\footnote{GitHub:~\href{https://github.com/RanyaJumah/EDGY}{https://github.com/RanyaJumah/EDGY}}).
\end{itemize}
%

\section{\uppercase{Disentanglement}}
\label{sec:disentang}

In this section, we provide a brief overview of the necessary technical background about disentanglement and its models.

\subsection{Learning Disentangled Representation}
There has been notable recent interest in learning disentangled representations in various domains, such as computer vision~\cite{hadap2020neural}, ML fairness~\cite{sarhan2020fairness, NIPS2019_8699}, and domain adaptation~\cite{tsai2019domain, peng2019domain}, as they promise to enhance robustness, interpretability, and generalization to unseen examples on downstream tasks. The overall goal of disentangling is to improve the quality of the latent representations by explicitly separating the underlying factors of the observed data~\cite{kim2018disentangling}. For example, in computer vision, there is a variety of tasks that have benefited from disentangled representations like pose-invariant recognition~\cite{reed2014learning}, attribute transfer via adversarial disentanglement~\cite{zhao2019look}, and person re-identification~\cite{eom2019learning}. 

There is an extended trend towards learning disentangled representations in the speech domain. Speech signal simultaneously encodes linguistically relevant information, e.g. phoneme and linguistically irrelevant information like speaker characteristics. In the case of speech processing, an ideal disentangled representation would be able to separate fine-grained factors such as speaker identity, noise, recording channels, and prosody~\cite{gong2018towards}, as well as the linguistic content. Thus, disentanglement will allow learning of salient and robust representations from the speech that are essential for applications including speech recognition~\cite{park2019unsupervised}, prosody transfer~\cite{sun2020fully, zhang2019learning}, speaker verification~\cite{peri2020empirical}, speech synthesis~\cite{sun2020fully,hu2020unsupervised}, and voice conversion~\cite{huang2020unsupervised}, among other applications.

\subsection{Disentanglement Models}
Most prior works on disentangled representation learning are based on well-established frameworks, such as variational autoencoders (VAEs)~\cite{kingma2013auto} and generative adversarial models (GANs)~\cite{goodfellow2014generative} in learning disentangled and hierarchical representations. They are based on the original objective of these models and derive regularizations to strengthen the disentanglement to learn compact and meaningful representations. These works can be categorized into three groups according to the model that depend on: VAE-based models~\cite{higgins2017beta, lample2017fader, NIPS2015_5851}, GAN-based models~\cite{chen2016infogan,choi2018stargan, 2017learning}, and combinations of AEs and GANs~\cite{lee2020high,makhzani2015adversarial,engel2017latent}. While extensive progress was made by these prior works in the computer vision domain, little has been done for speech processing.

Learning speech representations that are invariant to differences in speakers, language, environments, microphones, etc., are incredibly challenging to capture~\cite{latif2020deep}. To address this challenge, variants of VAEs have recently been proposed in learning robust disentangled representation owing to their generative nature and distribution learning abilities. Hsu et al., in~\cite{hsu2017unsupervised}, propose the Factorized Hierarchical VAE (FHVAE) model to learn hierarchical representation in sequential data such as speech at different time scales. Their model aims to separate between sequence-level and segment-level attributes to capture multi-scale factors in an unsupervised manner. Similarly, Predictive Aux-VAE~\cite{springenberg2019predictive} was proposed to obtain speech representations at different timescales by disentangling local (content) from global (speaker) information inherently. Although the focus of these works is to raise the efficiency and effectiveness of speech processing applications (e.g. speech recognition, speaker verification, and language translation), in this paper we highlight the benefit of \emph{learning disentangled representation} to learn privacy-preserving speech representations, as well as showing how disentanglement can be useful in transparently protecting user privacy.
\section{\uppercase{Problem Description}}
\label{sec:threatmodel}
In this section, we present our threat model and explain the goals of the user, the potential attribute inference attacker, and the \emph{defender} in this context.
\subsection{User}
Users provide information to cloud service providers to maximize their utility under the assumption that sensitive information in the data should be protected. They agree on the use of data for a specific task (i.e. execution of a voice command), but they may not consent to their data being used for additional analyses that may violate their privacy. In the voice control scenario, while users (data owners) may agree to share their voice recordings for speech recognition and accurate execution of their command, they might want to protect their sensitive information (e.g., emotion or health status) such that no secondary inferences are made from the data. For example, Amazon has patented technology to analyze users' voices to determine their emotions and/or mental health conditions. This allows a deeper insight into the user's mental state, which can be exploited to serve highly targeted content~\cite{AmazonPatents_2018}.

\subsection{Attacker}
Our attack aims to correctly infer sensitive attributes (e.g., gender, emotion, and health status) about data owners by exploiting a secondary use of the same data collected for the main task. Specifically, the attacker could be any party (e.g., a service provider, advertiser, data broker, or a surveillance agency) which has interest in data owners’ sensitive attributes. The service providers could use these attributes for targeting content; or data brokers might profit from selling these information to other parties such as advertisers and insurance companies, and surveillance agencies may use these attributes to recognize users and track their activities and behaviour. 
In this paper, we focus on the following question: to what extent can such an attacker infer data owners' sensitive attributes, and to what extent can this be prevented. To answer this, we assume that the attacker has white-box knowledge (i.e. parameters and target model architecture) and a machine learning classifier that uses data owners’ data as input to train the classifier and predict data owners' sensitive attributes.
 
\subsection{Defender}
\begin{figure}[t!]
  \centering
  \includegraphics[width=\columnwidth]{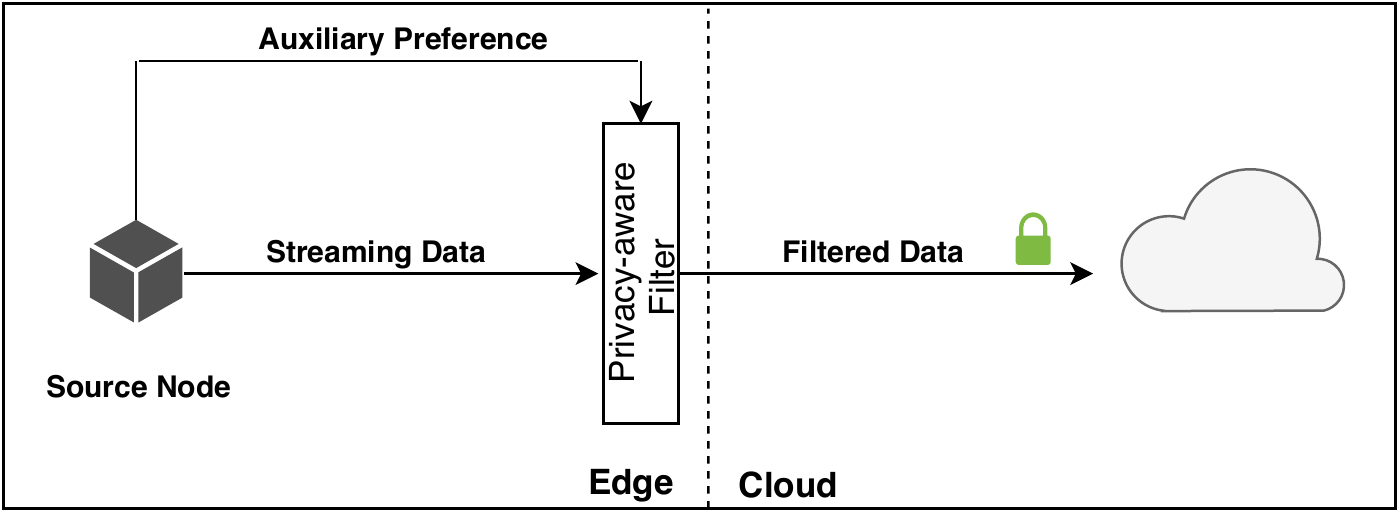}
  \caption{The workflow of the proposed framework: it serves as a filter between the edge and the cloud to purify data from a source node based on an auxiliary user preference
  }
      \label{fig:edge-cloud}
\end{figure}

The goal of the privacy-preserving framework in this paper is to protect the sensitive attributes of data shared against potential attribute inference attacks launched by a curious attacker. We propose a privacy-aware defense framework controlled by the data owner to filter the raw data at the edge before sharing it with cloud service providers, as shown in Figure~\ref{fig:edge-cloud}. The proposed framework works as a bridge between the data owners and the service providers to allow privacy-preserving communication between them. This framework receives the raw data as well as user preferences as auxiliary information, then it uses the user preference to filter (i.e. remove) sensitive attributes, which would be otherwise contained in their shared data.

Algorithm 1 gives the overall workflow of the proposed framework to reconstruct the filtered data $\mathrel{\bar{x}}$ using \emph{learning disentangled representation}. We call the proposed framework Dual-phase Disentangled Filter (DDF). Firstly, the DDF receives the inputs, which are raw data $x$, as well as user privacy preferences $P$ within one of the options provided by the DDF. To demonstrate the concept, we enable three levels: high, moderate, and low. The option is user configurable, and may change for differing application domains, service contexts, etc. A privacy preference $P$ is associated with the set of tasks, resulting in a list of tasks that can be performed on the raw data $x$. Phase II begins by checking the contents of the privacy preference $P$ list. In the case that it is empty, the user prefers to share data without filtering it. Otherwise, the raw data $x$ along with privacy preference $P$ list will be passed to the disentangle module, which starts different branches, each attempting to learn independent information related to a specific task. After the disentanglement, the decoder $D$ reconstructs the filtered data $\mathrel{\bar{x}}$ by receiving the concatenation of the output of the desired branches.

\section{\uppercase{Dual-phase Disentangled Filter}}
\label{sec:defend}
\begin{algorithm}
    \SetKwInOut{Input}{Input}
    \SetKwInOut{Output}{Output}
  \Input{Raw data x, Privacy preference $P$}
  \Output{Filtered data $\mathrel{\bar{x}}$}
  initialization\;
\SetKwFunction{FMain}{UserPreference}
      \SetKwProg{UPre}{Function}{:}{}
      \UPre{\FMain{}}
      { 
      \tcp{set the mode}
      \For{$i\gets1$ \KwTo $m$}
      {
        $P$ $\gets$ \{ [$t_{1}$; $t_{2}$; ... ; $t_{i}$]: task i belong to preference p \}
      }
        \KwRet $P$
      }
\SetKwFunction{FMain}{Encoder}
      \SetKwProg{Enc}{Function}{:}{}
      \Enc{\FMain{$x$}}
      {
        $z$ $\gets$ $fnn$ ($x$)\\
        \KwRet $z$
      }
\SetKwFunction{FMain}{Disentangle}
      \SetKwProg{Dis}{Function}{:}{\KwRet}
      \Dis{\FMain{$x$, $P$}}
      {
      \tcp{for each task $t_{i}$ in $P$}
      \For{$i\gets1$ \KwTo $m$}
        {
        \tcp{disentangled branch}
        $z$ $\gets$ Encoder(x)\\
        $B_{i}$ $\gets$ $fnn$ ($z$)\\
        $\mathrel{\bar{z}}$ $\gets$ [$B_{1}$; $B_{2}$; ... ; $B_{i}$] \tcp{concatenate}
        }
      \KwRet $\mathrel{\bar{z}}$
      }
\SetKwFunction{FMain}{Decoder}
      \SetKwProg{Dec}{Function}{:}{\KwRet}
      \Dec{\FMain{$\mathrel{\bar{z}}$}}
      {
        $\mathrel{\bar{x}}$ $\gets$ $fnn$ ($\mathrel{\bar{z}}$)\\
        \KwRet $\mathrel{\bar{x}}$
      }
$P$ $\gets$ UserPreference()\\
\eIf{$P$ not Empty}
    {
      $\mathrel{ \bar{z}}$ $\gets$ Disentangle($x$, $P$ )\\
      $\mathrel{\bar{x}}$ $\gets$ Decoder($\mathrel{\bar{z}}$)\\
    }
    {
      $\mathrel{\bar{x}}$ = x
    }
\KwRet $\mathrel{\bar{x}}$
\caption{Dual-phase Disentangled Filter}
\end{algorithm}

\subsection{Overview}
We focus on the setting where the users' preferences serve as a control signal over a utility-privacy optimization problem. The users' inclusion can enable them to manage their information flow and potentially make better decisions on sharing their data to reduce privacy concerns. However, the major challenge to adjust this setting is how to learn disentangled and robust representations from the users' input that reflect their privacy preference. To tackle this challenge, we propose a DDF framework that builds upon VAEs~\cite{kingma2013auto} to encourage learning these disentangled latent representations and then using users' preferences to filter out unwanted representations. This is inspired by recommender systems, where giving users explicit control over the filtering process can enhance explainability and transparency in sharing their data.

In Phase I (\textbf{Optimization}), we categorize users' preferences into $i$ options $P_{i}$, which may be based on the application domain (e.g. audio analysis). For each option $P_{i}$, there exists a set of tasks $T_{i}$ that are associated with it. When specifying a preference option $P_{i}$, the tasks associated with this preference will achieve high accuracy, while the rest of the tasks may have low accuracy. The relation between the preference option $P_{i}$ and the task $T_{i}$ is denoted by $y_{p,t}$ $\in$ $\{ 0, 1 \}$, where $y_{p,t}$ = 1 indicates that preference explicitly adopts task $i$, whereas $y_{p,t}$= 0 means there is no relation between the two. 

In Phase II (\textbf{Filtering}), we propose an autoencoder architecture with a disentangle module to explicitly decouple the distinct factors in the raw data. Firstly, the disentangle $R$, which is the key module in the proposed framework, receives a user's raw data $x$ and privacy preference $P_{i}$. Based on the preferred option, the disentangle $R$ starts a particular branch for each task $B_{t}$. Each branch $B_{t}$ aims to learn task-specific representations $r_{t,s}$, while ignoring task-invariant representations $r_{t,i}$. Then, the branches' outputs of the target tasks are concatenated to form a disentangle $R$ output $\mathrel{\bar{z}}$. Finally, the Decoder $D$ uses the disentangle $R$ output $\mathrel{\bar{z}}$ to reconstruct the filtered data $\mathrel{\bar{x}}$.

\subsection{DDF for Speech Representation}
\begin{figure}[t!]
  \centering
  \includegraphics[width=\columnwidth]{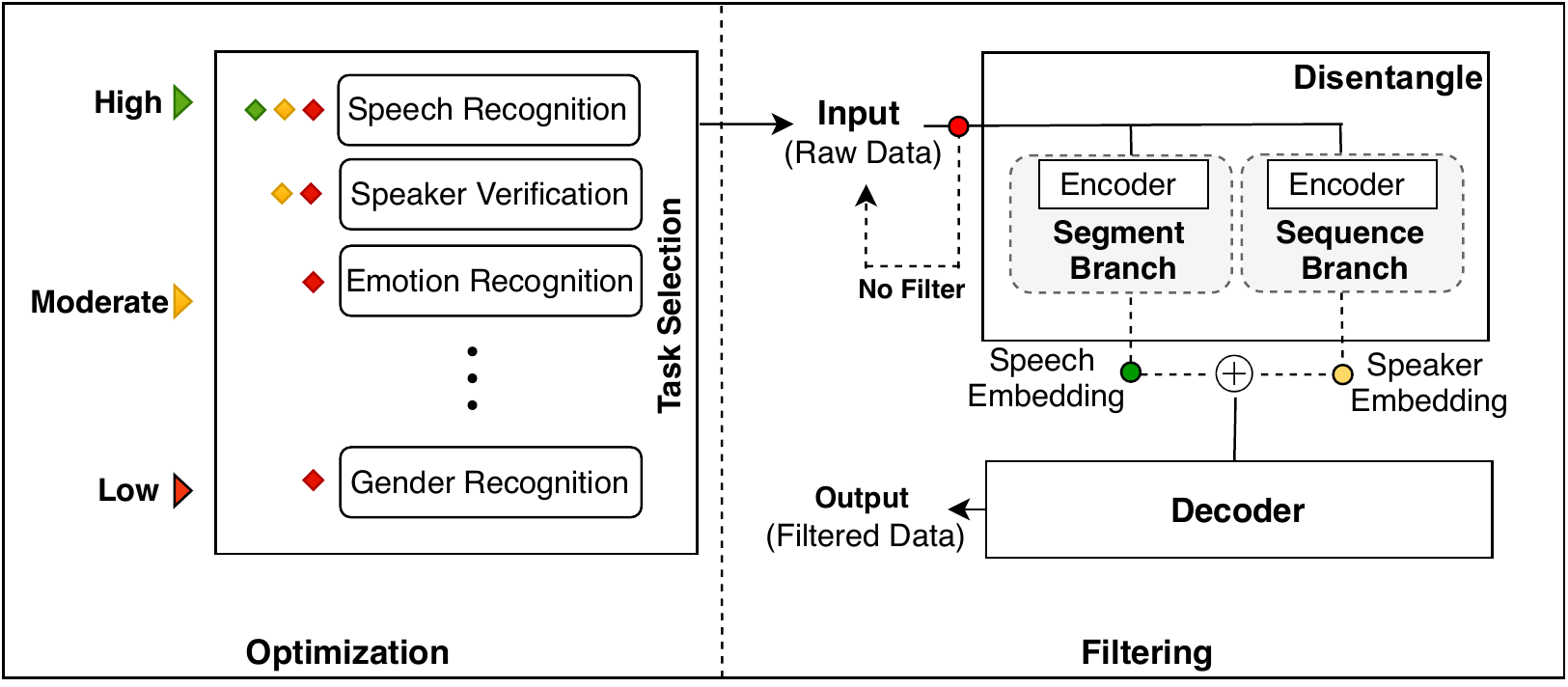}
  \caption{The proposed framework begins by adjusting the privacy preferences (high, moderate, and low; left) that are used as a control signal to extract the corresponding representations and reconstruct the output (right)}
      \label{fig:featuresComp}
\end{figure}

Leveraging the multi-scale nature of sequences such as speech, text, and video, distinct factors can be captured at different timescales \cite{hsu2017unsupervised}. For example, in speech signals the phonetic content affects the segment level, while the speaker characteristic affects the sequence level. Thus, the speech signal can be disentangled into several independent factors, each of which carries a different type of information. 
In our context, the idea is to \emph{disentangle} the factors related to the task we want to compute. We aim to demonstrate the effectiveness of \emph{learning disentangled representation} in preserving the sensitive attributes in the user data. This disentanglement can also be beneficial to promote transparency in protecting users' privacy. Figure~\ref{fig:featuresComp} illustrates our use of the \emph{disentangled representation} to enable users' control over the data they want to share.

\subsubsection{\textbf{Phase I}}
We consider three preference options: high $P_{h}$, moderate $P_{m}$ and low $P_{l}$. We also suppose there are three main tasks that can be performed on the user data: speech recognition $T_{1}$, speaker verification $T_{2}$, and others (later emotion and gender recognition) $T_{3}$. For each option $P_{i}$, we associate a set of tasks $T_{i}$. For example, when a user specifies a preference option $P_{h}$, the user's raw data $x$ will be used for the $T_{1}$, while the rest of the tasks $T_{2}$ and $T_{3}$ will get mistaken results. As the relation between the preference option $P_{i}$ and the task $T_{i}$ is denoted by $y_{p,t}$ $\in$ $\{ 0, 1 \}$, then $y_{h,1}$ = 1, whereas $y_{h,2}$ and $y_{h,3}$ = 0. Similarly, when the user selects a preference option $P_{m}$, then $y_{m,1}$ and $y_{m,2}$ = 1, whereas $y_{m,3}$ = 0. For the last preference option $P_{l}$, $y_{l,1} = y_{l,2} = y_{l,3} = 0$, which means no filter operation will be done over the user's raw data $x$. 

\subsubsection{\textbf{Phase II}}
Intuitively, autoencoders use an encoding network to extract a latent representation, which then passes through a decoding network to recover the original data. Autoencoders are trained to minimize the reconstruction error between the encoded-decoded data and the raw data. VAE is an autoencoder whose encodings distribution is regularized during training to ensure that its latent space captures useful representation to allow generating powerful new data. VAE consists of the following main parts: an encoder network for modelling a posterior distribution q($z$|$x$) of discrete latent random variables z given the input data x, a prior distribution p($z$), and a decoder with a distribution p($x$|$z$) over the input data. $VAE_{loss}$ decomposes into reconstruction loss of standard autoencoder and Kullback-Leibler (KL) divergence between the prior p($z$) and the posterior distribution q($z$|$x$). The joint minimization of both losses leads to reasonable reconstruction while reducing the latent space dimension at the same time.

In the speech domain, there are different variations of VAE that aim to learn \emph{disentangled representation}~\cite{hsu2017unsupervised,sun2020fully} to allow disentangling and controlling different attributes within the speech signal such as speech content, speaker identity, and emotion. Thus, to achieve our goal in \emph{learning disentangled representation} for privacy preservation purposes, we use different methods to obtain these representation. Details about the implementation of each module are as follows:

\textbf{Disentangle $R$}
We intend to disentangle speech representations from the input speech explicitly into several factors that can be used independently for different tasks. To achieve this, we divide the disentangle module into separate branches to force learning diverse types of information~\cite{mathieu2016disentangling}. We use a combination of objectives to encourage these different branches to learn task-related factors. Assuming we have two basic tasks, speech recognition and speaker verification, that we want to maintain, we have two branches to learn independent factors for each.

\textbf{Branch 1 ($B_{speech-recognition}$)}\\
Inspired by Vector Quantized VAE (VQ-VAE) in~\cite{van2017neural}, we perform vector quantization to extract the phonetic content while being invariant to low-level information. VQ-VAE model aims to produce discrete latent space using Vector Quantization (VQ) techniques. During the forward pass, the output of the encoder $z_{e}$(x) is mapped to the closest entry $c_{i}$ in a discrete codebook of $c$ = [$c_{1}$,$c_{2}$,..,$c_{k}$]. Precisely, VQ-VAE finds the nearest codebook using Eq.1 and uses it as the quantized representation $z_{q}$(x) = $c_{q}$(x) which is passed to the decoder as content information.
\begin{equation}
q(x) = argmin_i\left \| z_e(x) - c_i \right \|_2^2 
\end{equation}
The transition from $z_{e}$(x) to $z_{q}$(x) does not allow gradient backpropagation due to the argmin function, but uses a straight-through estimator~\cite{bengio2013estimating}. VQ-VAE is trained using a sum of three-loss terms (in Eq.2): the negative log-likelihood of the reconstruction, which uses the straight-through estimator to bring the gradient from the decoder to the encoder, and two VQ-related terms - the distance from each prototype to its assigned vectors and the commitment cost~\cite{van2017neural}.
\begin{equation}
\begin{split}
L = \log p(x | z_q(x)) + \left \| sg[z_e(x)] - c_q(x) \right \|_2^2 + \beta \left \| z_e(x) − sg[c_q(x)] \right \|_2^2
\end{split}
\end{equation}
sg(·) denotes the stop-gradient operation that zeros the gradient with respect to its argument during backward pass, which effectively constraining its operand to be a non-updated constant. The VQ within VAE replaces the continuous latent vectors with deterministically quantized one, and thus it will encourge learning latent representations with powerful disentanglement between the phonetic content and speaker identity. 

By using vector quantization as a regularizer, the encoder in this branch is encouraged to extract content-specific representations and discard the invariant representations that the decoder can infer from the information of the other branch for reconstruction purposes. Alternatively, we can use the output of this branch as speech embedding to train models that use these discrete representations directly to translate from speech to text instead of reconstruction, which may cause a significant improvement in privacy protection in sharing speech data, as shown in Figure~\ref{fig:featuresComp}. For example, similar to VQ-VAE~\cite{van2017neural}, vq-wav2vec was proposed by~\cite{Baevski2020vq-wav2vec} to quantize the dense representations from the speech segments by implementing either a Gumbel-Softmax or online k-means clustering. Then, they apply well-performing NLP algorithms (e.g. BERT) to these quantized representations and they present promising state-of-the-art results in phoneme classification and speech recognition. 

\textbf{Branch 2 ($B_{speaker-verification}$)}\\
Obtaining a good speaker representation becomes particularly important in speaker recognition, speaker adaptation, and other applications, where irrelevant information in the signal should be filtered out. Although speaker recognition systems can vary widely in their design, they share the same objective in finding discriminative representations to maintain high accuracy and robustness in a variety of environments. 

The goal of this branch is to learn such speaker representations that preserve user identity. To achieve this, we use two different methods to extract these representations. Firstly, we use a one-hot speaker code~\cite{hojo2018dnn} to extract the speaker's representations and then use this code as a global condition for the decoder to reconstruct the speech signal. Alternatively, we use Thin ResNet-34~\cite{xie2019utterance} trained using the angular variant learning metric~\cite{chung2020defence} to encourage learning discriminative representation. The encoder in this branch will encourage the extraction of speaker-specific representations and discard invariant representations the decoder can infer from information of the other branch for reconstruction. To support our goal of enhancing privacy protection in sharing speech data, we point out that the output of this branch can be used independently as a speaker embedding, as shown in Figure~\ref{fig:featuresComp}, for speaker verification application instead of reconstructing.

\begin{figure}[t!]
  \centering
  \includegraphics[width=\columnwidth]{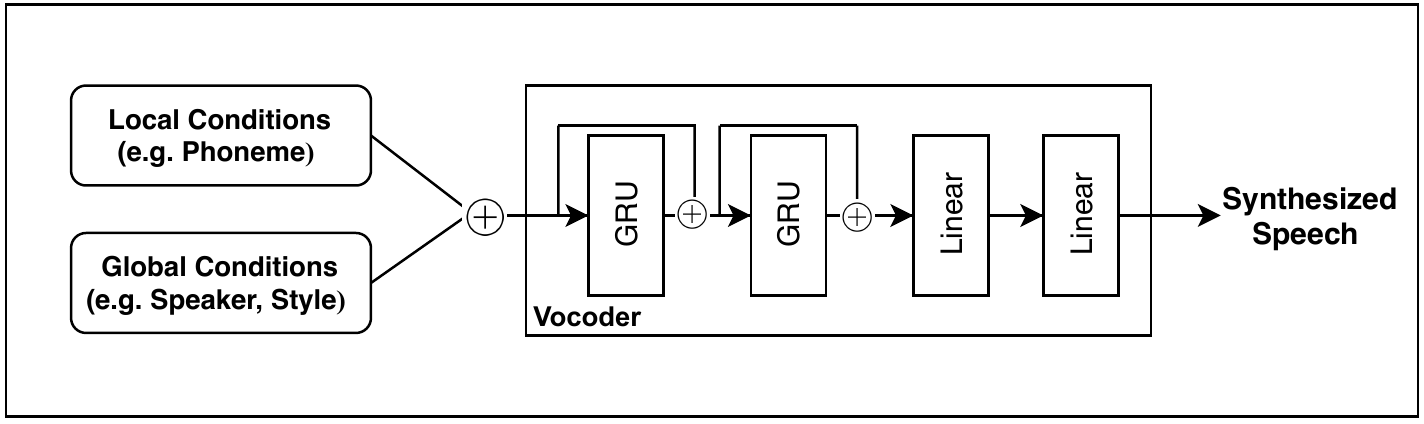}
  \caption{An overview of the Vocoder's workflow: it concatenates a global (sequence) and local (segment) to reconstruct the output (WaveRNN~\cite{kalchbrenner2018efficient})}
      \label{fig:decoder}
\end{figure}

\textbf{Decoder $D$}
In the speech domain, a \emph{vocoder} learns to reconstruct audio waveforms from acoustic features~\cite{oord2016wavenet}, as shown in Fig.~\ref{fig:decoder}. Traditionally, the waveform can be vocoded from these acoustic or linguistic features using handcrafted models such as WORLD~\cite{morise2016world}, Straight~\cite{kawahara2006straight}, and Griffin-Lim~\cite{griffin1984signal}. However, the quality of those traditional vocoders was limited by the difficulty in accurately estimating the acoustic features from the speech signal.

Neural vocoders such as Wavenet~\cite{oord2016wavenet} have rapidly become the most commonly used vocoding method for speech synthesis. Although it improved the quality of generated speech, it has significant cost in computation power and data sources, and suffers from poor generalization~\cite{lorenzo2018towards}. To solve this problem, many architectures such as Wave Recurrent Neural Networks (WaveRNN)~\cite{kalchbrenner2018efficient} have been proposed. WaveRNN combines linear prediction with recurrent neural networks to synthesize neural audio much faster than other neural synthesizers. 
In our framework, we use WaveRNN as a decoder with a minor change suggested by~\cite{lorenzo2018towards}. The autoregressive component consists of a single forward gated recurrent unit (GRU) (hidden size of 896) and a pair of affine layers followed by a softmax layer with 1024 outputs, predicting the 10-bit mu-law samples for a 24 kHz sampling rate. The conditioning network consists of a pair of bi-directional GRUs with a hidden size of 128. The autoregressive component captures the content, while the conditioning component represents the speaker's characteristics. To achieve our goal of preserving privacy, the quality of generated speech is measured by the extent to which it contains the desired information after the filtering process and removing invariant information. 

In general, Phase II is designed by taking advantage of the disentanglement in learning independent representations from the input, and then Phase I output is used to determine the outputs of the proposed framework. Phase II is intended to accommodate preferences assuming that the input is passed on several branches to learn different information according to the specific task of the branch.

\section{\uppercase{Experiments}}
\label{sec:experments}

In this section, we describe the datasets, inference attack models, and proposed framework settings. We conduct our experiments on a Z8 G4 workstation with Intel (R) Xeon (R) Gold 6148 (2.8 GHz) CPU and 256~GB RAM. The operating system is Ubuntu 18.04. We train all the models using PyTorch~\cite{PyTorch} on an NVIDIA Quadro RTX 5000 GPU. 
\subsection{Datasets}
We use five real-world datasets recorded for various purposes such as speech recognition, speaker recognition, and emotion recognition. The details of each dataset are as follows:\\
\textbf{IEMOCAP.} The Interactive Emotional Dyadic Motion Capture dataset~\cite{busso2008iemocap} has 12 hours of audio-visual data from 10 actors where the recordings follow dialogues between a male and a female actor in both scripted or improvised topics in the English language. The data was segmented by speaker turn, resulting in 5,255 scripted recordings and 4,784 improvised recordings. It was mainly recorded to facilitate the development of multimodal emotion recognition systems. We use the scripted recordings that were labeled with four emotions: anger, happy, sad, and neutral.
\newline
\textbf{RAVDESS.}
The Ryerson Audio-Visual Database of Emotional Speech and Song~\cite{RAVDESS} contains 1,440 recording for 24 actors (12 male and 12 female), vocalizing two lexically-matched statements in a neutral North American accent. It was recorded to facilitate the development of multimodal emotion recognition systems. It includes seven emotions: calm, happy, sad, angry, fearful, surprise, and disgust, as well as neutral expression. We use the entire dataset. 
\newline
\textbf{SAVEE.} Surrey Audio-Visual Expressed Emotion database~\cite{haq2008audio}. It consists of phonetically-balanced sentences from standard TIMIT (acoustic-phonetic continuous speech dataset) uttered by four English actors with a total size of 480 utterances. It was primarily recorded to facilitate the development of multimodal emotion recognition systems. It contains expressions of seven emotions: calm, happy, sad, angry, fearful, surprise, and disgust, as well as neutral. We use the entire dataset. 
\newline
\textbf{LibriSpeech.}
LibriSpeech~\cite{panayotov2015librispeech} is a large dataset of approximately 1,000 hours of reading of English. It was derived from reading audiobooks from the LibriVox project, and was recorded to facilitate the development of automatic speech recognition systems. We use the train-clean100 set. 
\newline
\textbf{VoxCeleb.}
The VoxCeleb dataset~\cite{nagrani2017voxceleb} contains over 100,000 utterances for 1,251 celebrities, extracted from videos uploaded to YouTube. It was curated to facilitate the development of automatic speaker recognition systems. We use the VoxCeleb2 subset of about 1,200 recordings. 
\newline
\textbf{Training and Testing.} We divide each dataset using 80$\%$ for training and 20$\%$ for testing. 

\subsection{Attribute Inference Attacks}

\begin{table*}[]
\Large
\caption{Accuracy of attribute inference attack using different acoustic models to extract the representation (G=gender (binary); E=emotion)}
\label{tab:inference_accuracy}
\begin{adjustbox}{width=\textwidth}
\begin{tabular}{c|c|c|c|c|c|c|c|c|c|c|c|c|c|c|}
\cline{2-15}
\multicolumn{1}{l|}{\textbf{}} & \multicolumn{7}{c|}{\textbf{wav2vec Model}} & \multicolumn{7}{c|}{\textbf{DeepSpeech2 Model}} \\ \hline
\multicolumn{1}{|c|}{\multirow{2}{*}{\textbf{\begin{tabular}[c]{@{}c@{}}Attacker \\ Model\end{tabular}}}} & LibriSpeech & VoxCeleb & SAVEE & \multicolumn{2}{c|}{IEMOCAP} & \multicolumn{2}{c|}{RAVDESS} & LibriSpeech & VoxCeleb & SAVEE & \multicolumn{2}{c|}{IEMOCAP} & \multicolumn{2}{c|}{RAVDESS} \\ \cline{2-15} 
\multicolumn{1}{|c|}{} & G(\%) & G(\%) & E(\%) & G(\%) & E(\%) & G(\%) & E(\%) & G(\%) & G(\%) & E(\%) & G(\%) & E(\%) & G(\%) & E(\%) \\ \hline
\multicolumn{1}{|c|}{LR} & 85.8 & 90.4 & 62.2 & 82.9 & 56.4 & 99.4 & 74.4 & 60 & 78.3 & 53.1 & 58.8 & 47.7 & 93 & 57.2 \\ \hline
\multicolumn{1}{|c|}{RF} & 86.7 & 80.8 & 43.2 & 86.4 & 55 & 95.6 & 61.9 & 50.7 & 63.5 & 42.2 & 62 & 50.1 & 86 & 53.5 \\ \hline
\multicolumn{1}{|c|}{MLP} & 75.8 & 78.8 & 39 & 76.4 & 51.2 & 93.8 & 64.4 & 56.7 & 57.8 & 40.5 & 58.4 & 45.3 & 95.3 & 63.2 \\ \hline
\multicolumn{1}{|c|}{SVM} & 76.7 & 85.6 & 55.7 & 85 & 57.9 & 94.4 & 60.2 & 66.7 & 73.9 & 46.2 & 54.3 & 55.6 & 88.4 & 61 \\ \hline
\end{tabular}
\end{adjustbox}
\end{table*}

An attribute inference attack aims to infer sensitive information from users' recordings. Specifically, an attacker trains a particular classifier that takes the representation extracted from users' recordings as input and infers sensitive attributes (e.g., emotion and gender).

\subsubsection{Target Attributes.}
~We test the proposed framework over binary (i.e., gender) and non-binary (i.e., emotion) attributes. For IEMOCAP and RAVDESS, we consider inference tasks are emotion recognition and binary gender attributes, and train separate models to classify emotion and gender recognition for the entire representation (after extracting these representations from the raw recording) for each dataset. For LibriSpeech and VoxCeleb, we consider the inference task to be gender, and we train separate models to classify gender for the entire representation for each dataset. For SAVEE, as it contains one gender, we only consider the emotion inference. We repeat this setting for each type of attacker classifier (35 models in total).

\subsubsection{Models.} ~Below are the details for each attack classifier:

\textbf{Logistic Regression (LR):}  LR is a machine learning classification algorithm used to predict the probability of a categorical dependent variable. For binary classification such as gender recognition, we use a sigmoid function to predict the true label, i.e. male or female based on a given representation. For multiclass prediction, we use the softmax function instead of the sigmoid function to normalize the input values from all classes between 0 and 1 and return the probabilities of each class. All models train using a stochastic average gradient (SAG) and for 300 iterations. In this attack, the attacker uses a LR classifier to perform attribute inference attacks. 

\textbf{Random Forest (RF):} RF is a machine learning classification algorithm that creates decision trees on randomly selected data samples, gets a prediction from each tree, and selects the best solution by the means of voting. All models implement 100 estimators, which indicates the number of trees in the forest. In this attack, the attacker uses a RF classifier to perform attribute inference attacks.

\textbf{Support Vector Machine (SVM):} SVM is a discriminative classifier to find a hyperplane in N-dimensional space (N: features numbers) that accurately classifies the data points. All models implement a radial basis function (RBF) as a kernel function to scale properly on large numbers of features in the input space, and scale gamma distribution. In this attack, the attacker uses SVM classifier to perform attribute inference attacks.

\textbf{Neural Network (NN)- Multilayer Perceptron (MLP):} In this attack, the attacker uses a three-layer fully connected neural network (input layer, a hidden layer which has 2048 neurons, and output layer) to perform attribute inference attacks. All models adopt the rectified linear unit (ReLU) as an activation function. They train by using Adam optimizer with learning rate = 0.001 and batch size = 200 for 300 iterations. As it is difficult to determine the possible structure of NNs, we chose a simple structure expected to be enough to analyze the captured information in extracted representations.

\subsubsection{Setup.}
~In advance of training these models, we must first extract the representations from various datasets using pre-trained acoustic models for speech recognition tasks. 
We extract the representations from raw audio in different datasets using wav2vec model~\cite{schneider2019wav2vec}, which achieves 2.43 $\%$ word error rate (WER) for speech recognition. The wav2vec relies on a fully convolutional architecture by applying two networks. The encoder network embeds the audio signal in a latent space and the context network combines multiple time-steps of the encoder to obtain contextualized representations. We use the pre-training model on the full 960-hour Librispeech training set with 32.5M parameters. To achieve our purpose of obtaining similar representations to those which may be used in acoustic models, we used only the output from the encoder network. The encoder layers have kernel sizes (10, 8, 4, 4, 4) and strides (5, 4, 2, 2, 2). The output of the encoder is a low-frequency feature representation $z_{i}$ $\in$ $Z$ that encodes about 30 ms of 16 kHz of raw audio and the striding results in representations $z_{i}$ every 10~ms. We then used these representations to train attacker classifiers. 
We extract the speech representation using state-of-the-art model DeepSpeech2~\cite{pmlr-v48-amodei16}, which reported a 6.71\% WER. It consists of 11 layers including bidirectional recurrent and convolutional layers. The model was trained using the CTC loss function and with a Stochastic Gradient Descent (SGD) and Momentum optimizer that was extended with the Layer-wise Adaptive Rate Clipping (LARC) algorithm. We use the pre-trained model to extract the feature representation $z_{i}$ $\in$ $Z$ from the log-spectrogram of the raw audio waveform signal. Then we used these representations (i.e., extracted using wav2vec and DeepSpeech2) to train attacker classifiers.

\subsection{Dual-phase Disentangled Filter Setting}
Firstly, spectrograms are generated from the raw time-domain waveform sampled at 16 kHz in a sliding window fashion using a Hamming window of width 25~ms and step 10~ms. For the speech embedding branch, these spectrograms are encoded by the encoder which consists of five residual convnet layers (using 768 units and ReLU activation). The encoder output (latent vectors) then passes through vector quantization (512 codebook size) to become a sequence of quantized representation that serves as the speech embedding. For the speaker embedding branch, the generated spectrograms are used as input for the encoder (Thin ResNet-34~\cite{xie2019utterance}) which is the same as the original ResNet with 34 layers, except cutting down the number of channels in each residual block to reduce computational cost. Self-attentive pooling (SAP)~\cite{cai2018exploring} is then used to aggregate frame-level features into the utterance-level representation that serves as speaker embedding. The representations from different branches are then upsampled and concatenated (using `torch.cat') to form the conditioning input to the WaveRNN decoder (Note: a one-hot vector representing the speaker can be used as a global condition of WaveRNN decoder).  We train the proposed framework on LibriSpeech, which has multiple speakers and was recorded at a sampling rate of 16 kHz. We used the Adam optimizer with an initial learning rate 4e-4 and evaluate the performance after 250,000 steps with batch size 64 (600,000 steps in total).

\section{\uppercase{Evaluation}}
\label{sec:evaluation}

In this section we evaluate our results in terms of (i) the effectiveness of the attributes inference attacks in voice processing using different model architectures and several datasets; and (ii) the efficiency of the proposed framework to defend against this class of attack in the voice domain.

\subsection{Attack Effectiveness}
\label{sec:evaluation.AE}
\subsubsection{Inference Accuracy.}
~Since the attacker's goal is to infer the target attribute, we evaluate an attack using the inference accuracy of the classifier used by the attacker. Precisely, we mean the accuracy of the classifier to infer sensitive information from the test set over the probability of the random guessing. Assuming, for example, that the sensitive attribute in question is the user's emotion, we have seven labelled categories in the available datasets (Ravdess and SAVEE). The random guess rate for success is therefore around 14\%. If we assume that the sensitive attribute is `gender' (e.g. binary male or female), the random guess rate will be 50\%. As the models potentially available to the attacker are unknown to us, we measure the success accuracy of various models to infer the target attribute trained on various datasets.

From Table~\ref{tab:inference_accuracy}, we see that the inference models have varying performance, ranging from about 40\% to 99.4\% in successfully inferring different attributes. This means that the inference attacks can improve accuracy by three or four times better over a random guess. The difference between these percentages reflects the extent to which the attributes relate to each other. For example, gender is more entangled with a speaker's identity than emotion, thus the attacker's success rate is higher in identifying speaker gender. Table~\ref{tab:gender} shows that although there is a reduction in the success rate of an attacker in identity speaker gender, still there is a slight increase over random guessing in some cases.

Moreover, the diversity in the datasets recorded in different environmental conditions and for diverse purposes may mimic the differences in the real-time environments for the deployment of voice-controlled devices. We notice that this diversity affects the attack success accuracy; shown in Table~\ref{tab:inference_accuracy}. For example, an attacker's success accuracy to infer emotion attribute is varied among the three emotional datasets (IEMOCAP, RAVDESS, and SAVEE), and the inference accuracy over RAVDESS is better than the other datasets due to the good quality of the emotional recordings. Despite these differences, we demonstrate that the deep acoustic models can be exposed to sensitive attributes extraction from their inputs.

\subsubsection{Impact of Acoustic Model Architecture on Attack Success.}
~We observe that the difference in the architecture of acoustic models can help attackers to successfully achieve their objectives. Insofar as the accuracy in extracting deep representations is increased to raise the efficiency of the speech processing tasks, the success percentage in inference of sensitive representations will also increase. For example, wav2vec~\cite{schneider2019wav2vec} has been developed to extract more powerful representations for speech recognition compared to the DeepSpeech2~\cite{pmlr-v48-amodei16} model. From Table~\ref{tab:inference_accuracy} we can see that the extracted representation using wav2vec increases the probability of the attacker inferring sensitive attributes compared with the DeepSpeech2 model. 

\begin{table}[]
\Small
\vspace{2.5mm}
\caption{Speech recognition and speaker verification measurements for voices  generated by the proposed framework with different privacy settings
}
\label{tab:WEREER}
\begin{adjustbox}{width=\columnwidth}
\begin{tabular}{c|c|c|c|c|}
\cline{2-5}
\multicolumn{1}{l|}{} & \multicolumn{2}{c|}{\textbf{\begin{tabular}[c]{@{}c@{}}Generated\\    (Hide Identity)\end{tabular}}} & \multicolumn{2}{c|}{\textbf{\begin{tabular}[c]{@{}c@{}}Generated\\ (Preserve Identity)\end{tabular}}} \\ \hline
\multicolumn{1}{|c|}{\textbf{Dataset}} & WER (\%) & EER (\%) & WER (\%) & EER (\%) \\ \hline
\multicolumn{1}{|c|}{LibriSpeech} & 1.16 & N/A & 0.32 & 0.03 \\ \hline
\multicolumn{1}{|c|}{VoxCeleb} & 0.80 & N/A & 0.13 & 0.0 \\ \hline
\multicolumn{1}{|c|}{IEMOCAP} & 0.86 & N/A & 0.29 & 0.07 \\ \hline
\multicolumn{1}{|c|}{RAVDESS} & 0.63 & N/A & 0.14 & 0.0 \\ \hline
\multicolumn{1}{|c|}{SAVEE} & 0.66 & N/A & 0.20 & 0.01 \\ \hline
\end{tabular}
\end{adjustbox}
\end{table}
\subsection{Defense Efficiency} 
\begin{table*}[]
\caption{Success accuracy in inferring the sex attribute after implementing the DDF framework with different privacy preference options (W2V: wav2vector model, DS: DeepSpeech2 model, Mod.:moderate, Rec\_m: reconstructed speech with moderate option, Rec\_h: reconstructed speech with high option)}
\label{tab:gender}
\begin{adjustbox}{width=\textwidth}
\begin{tabular}{c|c|c|c|c|c|c|c|c|c|c|c|c|c|c|c|c|}
\cline{2-17}
\multicolumn{1}{l|}{} & \multicolumn{4}{c|}{LibriSpeech (\%)} & \multicolumn{4}{c|}{VoxCeleb  (\%)} & \multicolumn{4}{c|}{IEMOCAP (\%)} & \multicolumn{4}{c|}{RAVDESS (\%)} \\ \cline{2-17} 
 & \multicolumn{2}{c|}{Low} & Mod. & High & \multicolumn{2}{c|}{Low} & Mod. & High & \multicolumn{2}{c|}{Low} & Mod. & High & \multicolumn{2}{c|}{Low} & Mod. & High \\ \hline
\multicolumn{1}{|c|}{\begin{tabular}[c]{@{}c@{}}Attack\\ Model\end{tabular}} & \begin{tabular}[c]{@{}c@{}}Raw\\ (w2v)\end{tabular} & \begin{tabular}[c]{@{}c@{}}Raw\\ (DS)\end{tabular} & Rec\_m & Rec\_h & \begin{tabular}[c]{@{}c@{}}Raw\\ (w2v)\end{tabular} & \begin{tabular}[c]{@{}c@{}}Raw\\ (DS)\end{tabular} & Rec\_m & Rec\_h & \begin{tabular}[c]{@{}c@{}}Raw\\ (w2v)\end{tabular} & \begin{tabular}[c]{@{}c@{}}Raw\\ (DS)\end{tabular} & Rec\_m & Rec\_h & \begin{tabular}[c]{@{}c@{}}Raw\\ (w2v)\end{tabular} & \begin{tabular}[c]{@{}c@{}}Raw\\ (DS)\end{tabular} & Rec\_m & Rec\_h \\ \hline
\multicolumn{1}{|c|}{LR} & 85.8 & 60 & 53.8 & 43.8 & 90.4 & 78.3 & 57.1 & 54.0 & 82.9 & 58.8 & 55.7 & 41.5 & 99.4 & 93 & 69.1 & 48.2 \\ \hline
\multicolumn{1}{|c|}{RF} & 86.7 & 50.7 & 55.0 & 46.6 & 80.8 & 63.5 & 64.2 & 52.3 & 86.4 & 62.2 & 57.4 & 48.7 & 95.6 & 86 & 53.4 & 49.2 \\ \hline
\multicolumn{1}{|c|}{MLP} & 75.8 & 56.7 & 52.7 & 46.9 & 78.8 & 57.8 & 51.1 & 42.2 & 76.4 & 58.4 & 60.0 & 44.9 & 93.8 & 95.3 & 67.4 & 41.7 \\ \hline
\multicolumn{1}{|c|}{SVM} & 76.7 & 66.7 & 60.2 & 54.3 & 85.6 & 73.9 & 62.2 & 49.7 & 85 & 54.3 & 66.2 & 47.1 & 94.4 & 88.4 & 55.9 & 45.6 \\ \hline
\end{tabular}
\end{adjustbox}
\end{table*}
\begin{figure*}[t!]
  \centering
  \includegraphics[width=\textwidth]{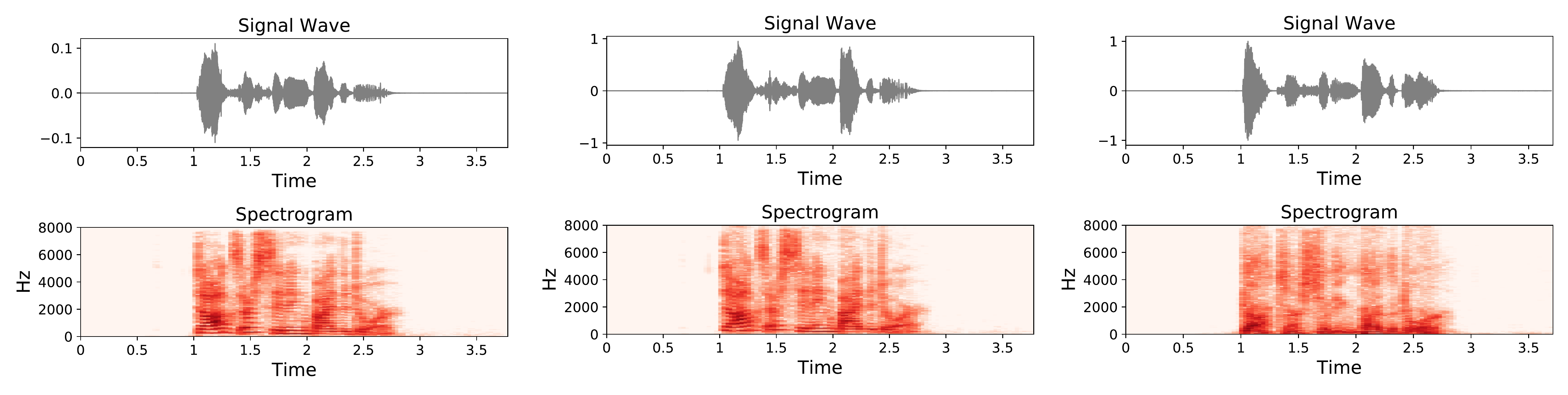}
  \caption{Spectrogram analysis for (left) raw speech, (middle) reconstructed speech preserved speaker identity, and (right) reconstructed speech with synthetic identity that contains the same speech content\\ }
      \label{fig:specto}
\end{figure*}

\subsubsection{Disentanglement and Controllability.}
~We aim to enable users to have control over their data by taking advantage of disentangled representation learning. Thus, we design and implement the proposed framework for three privacy preferences options, namely high, moderate, and low. After training the framework to explicitly learn the disentangled representation from the speech data, it can generate different outputs that reflect the selected privacy preferences. Setting the `high' option, speech content representation will be disentangled from the speaker's identity. 

The proposed framework can generate two types of output, either speech embedding or reconstruction of speech by concatenating the speech embedding with a synthetic identity. For the moderate option, the proposed framework can generate three types of outputs, which are speech embedding, speaker embedding, or reconstruction of speech by preserving the identity of the speaker while filtering out other information (e.g, emotions). Finally, by selecting the `low' option, the proposed framework will send the raw data without any filtering. Figure~\ref{fig:specto} shows the spectrogram of the reconstructed speech signal for the different options. The reconstruction recordings have the same content (same text), but the waveform is different and the prosody in the voice is modified. Moreover, we use word error rate (WER), a common metric of speech recognition performance, to use the difference in the word level between two spoken sequences to measure the difference in speech recognition between the raw speech signal and the reconstructed one for the different privacy preference options. We find, as shown in Table~\ref{tab:WEREER}, that there is an insignificant decrease in ($\sim$1\%) in speech recognition accuracy. Speaker verification is an example of a biometric system, where equal error rate (EER) is one measurement to predetermine the threshold values for its false acceptance rate (the ratio of the number of false acceptances divided by the number of identification attempts) and its false rejection rate (ratio of the number of false rejections divided by the number of identification attempts). In case the rates are equal, the common value is referred to as the equal error rate. The lower the equal error rate value will be, the higher the speaker verification accuracy. We use the EER to measure speaker verification accuracy (for the moderate privacy preference), and we find that an almost negligible rate between the raw and reconstructed speech signals for this speaker verification task. For `high' option, we did not test the EER for this option since the original speaker identity is discarded (i.e., only protect the speech content), and we use mimic identity to reconstruct the speech which contains the same content but different speaker identity.

Learning these disentangled representations not only serves our purpose to protect user privacy, but also is useful in finding robust representations for different speech processing tasks with limited data in the speech domain~\cite{latif2020deep}. 

\subsubsection{Privacy Estimation.}
\begin{figure*}[t!]
  \centering
  \includegraphics[width=\textwidth]{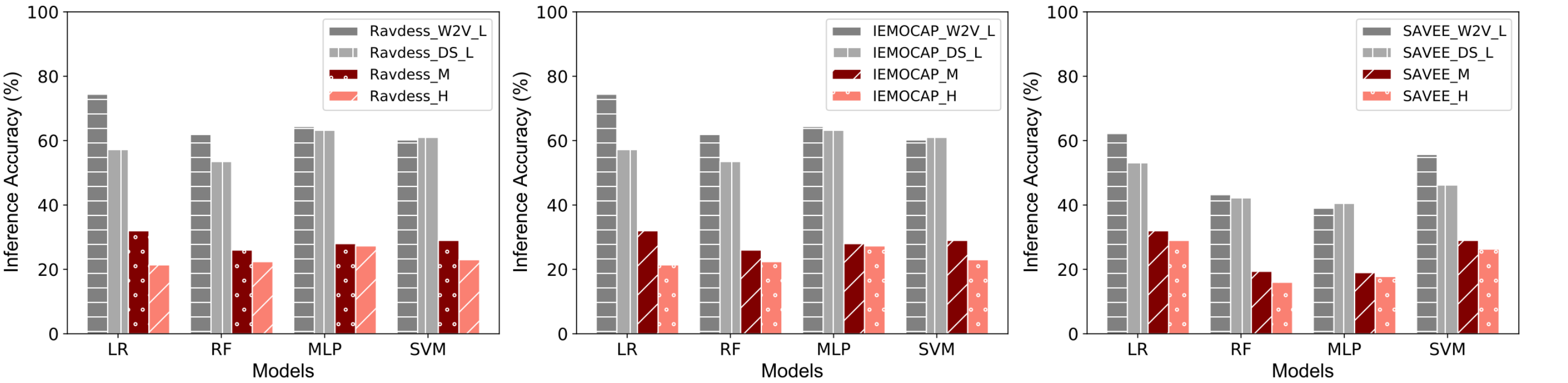}
  \caption{Accuracy in inferring the emotion attribute after implementing the DDF framework with diffrent privacy preference options (W2V: wav2vector model, DS: DeepSpeech2 model, L: low option, M: moderate option, and H: high option)}
      \label{fig:inference_accuracy}
\end{figure*}

~The \textbf{baseline} is the inference success from unfiltered representations. \\
\textbf{Privacy Preference: High}.
The output of the framework should reflect this privacy preference by achieving high accuracy in speech recognition while hiding a speaker's identity. Therefore, we measure the efficiency of the framework to learn a disentangled representation that preserves the speech content and discards the invariant information (i.e. speaker identity, emotion and gender) by examining an attacker's success in obtaining sensitive information using this representation. For fair comparison with the baseline inference accuracy, we only use the quantized embedding before concatenating it with a synthetic identity during reconstruction. Figure~\ref{fig:inference_accuracy} shows a considerable drop in the inference accuracy after implementing vector quantization (one technique) to learn such disentangled representations~\cite{van2017neural}, where the outcome is shown to be in line with guessing at random for all attacker models. 

\textbf{Privacy Preference: Moderate},
The output of the framework should reflect this preference by achieving high accuracy in speech recognition while preserving the speaker's identity. Thus, we measure the efficiency of the framework to learn a disentangled representation that preserves the speech content and speaker identity, and discards the invariant information (i.e. emotion and gender) by examining an attacker's success in obtaining this sensitive information using the output for this preference. Figure~\ref{fig:inference_accuracy} shows a notable reduction in the inference attacks' accuracy after reconstruction. This can be considered as a marginal improvement on random guessing. When comparing the results in Figure~\ref{fig:inference_accuracy}, we see that the speakers' representations may still preserve representations related to some sensitive attribute based on the slight rise in attacker success rate in emotion recognition. We also notice that the accuracy of gender recognition is higher in some cases (e.g., RF applied to LibriSpeech and MLP and SVM applied to IEMOCAP) and even compared to emotion recognition, which means gender closely related to the speaker's identity representation (i.e. highly related representation), as shown in Table~\ref{tab:gender}. In future work, we will investigate further disentanglement approaches (e.g., adversarial learning) within speaker embedding and add constraints as appropriate to try to limit this success. This could also be used to address various models used in different speech processing applications for extracting acoustic features from raw signals outperforming one another, e.g. wav2vector (self-supervised) outperforms DeepSpeech2 (supervised) in speech recognition, as mentioned in Sec.6.1.2.


\subsubsection{Prosody Visualization.}~Chroma feature (chromagram) is a fast and robust way to visualize audio attributes, and is relatively invariant to changes in the vocal tract resonances~\cite{wakefield1999chromagram}. This feature shows the distribution of energy along with the twelve different pitches or pitch classes, which refer to tones that share the same pitch-space (refers to tones sounding the same but separated by relative highness or lowness). To compute this feature, the spectrum is firstly computed in the logarithmic scale, with a selection of the 20 highest dB and restriction to a certain frequency range that includes an integer number of octaves. Then, the spectrum energy is redistributed along with the different pitches (i.e., chromas).  

\begin{figure*}[ht!]
  \centering
  \includegraphics[width=\textwidth]{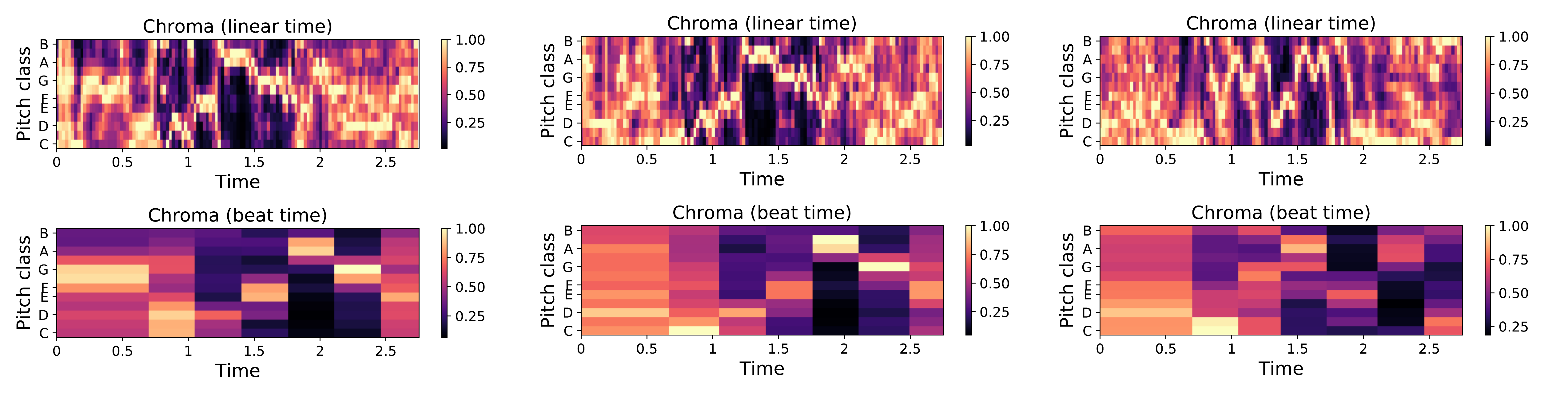}
  \caption{Chromagram analysis to measure the change in the prosodic features for emotion recognition for raw speech (left), reconstructed speech preserving speaker identity (middle), and reconstructed speech with synthetic identity (right)}
      \label{fig:chromo}
\end{figure*}

Prosodic features, like pitch, play an essential role in the transmission of vocal emotions~\cite{bulut2008robustness}. We therefore use chromagram visualization to measure the characteristics of the prosodic features between the raw speech and the reconstructed one. Figure~\ref{fig:chromo} compares the raw speech (angry emotion), the reconstructed speech with identity preserved (calm emotion), and the reconstructed speech with suppressed identity. It is clear that the change in the energy located in each pitch class for each frame reflects the success of the proposed framework in changing the prosodic representation related to the user’s emotion to maintain his/her privacy.

\section{\uppercase{Discussion and Future work}}
\label{sec:discussion_futurework}
Protecting users’ privacy where speech analysis is concerned continues to be a particularly challenging task. Yet, our experiments and findings indicate that it is possible to achieve a fair level of privacy while maintaining a high level of functionality for speech-based systems. Our results can be extended to shed light on several other questions discussed in this section.

\textbf{To what extent can speech representation be private?}
Our experimental evaluation highlights the vulnerability of the underlying acoustic models used by the speech processing systems (e.g. ASR systems) to potential attribute inference attacks. We estimate an attacker's success by running various arbitrary classifiers to measure the extent to which sensitive information can be obtained from a user's speech data. Based on the results shown in Table~\ref{tab:inference_accuracy}, we find that such an attacker has the opportunity to extract this information with a much higher degree of accuracy than would otherwise be possible by chance. For example, for emotion recognition using the RAVDESS dataset, and assuming that we have seven different emotions, then the random assumption rate will be $\sim$14\% of the time, but when using the logistic model the success rate is four times greater than this. When using the SVM model, i.e. a suggested model for analyzing emotions and physical conditions based on the Amazon patent~\cite{AmazonPatents_2018}, we observe that its success rate exceeds random guessing by three times. Although these classifiers are not ideal and the attackers can improve their strength by using more robust models (e.g. adversarial classifiers), our work aims to demonstrate these vulnerabilities and raise the alarm concerning the need for on-device solutions to sanitize user inputs insofar as possible before sharing them with service providers. 

\textbf{Is a two-phase framework necessary?}
The controllability enabled by the disentangled representations can help to design new privacy-preserving applications considering users' privacy preferences. This controllability will allow us to explicitly adjust the disentangled representation to match user privacy preferences. We expect that there are likely different user privacy preferences for analytics depending on the service providers with which they interact. For example, when users communicate with health service providers, they may prefer to share raw data without any filtering due to the urgent need to provide accurate information to trusted specialists. To accommodate such differences, we design a two-phase framework where the first phase captures user preferences, while the second phase learns disentangled representations to reflect these preferences.

As a first step, we have shown three privacy preference options (i.e. high, moderate, and low). Supposing that the user wants to interact with a smart home assistant such as Amazon’s Alexa or Google Home, for the high privacy preference option, the default analysis task should be to understand the user command and response based on it without any additional information that allows secondary processing or re-purposing of the user data. For the moderate privacy preference, the default analysis tasks should be speech-to-text and speaker recognition for authentication purposes, whereas the low privacy option allows users to share their data without any alteration. These are, however, just some examples of potential preferences and many more could be developed. In future work, we intend to provide users with additional privacy controls depending on the devices and services with which they are interacting.

\textbf{Is disentanglement necessary?}
Speech data has complex distributions and contains crucial information beyond linguistic content that may include information contained in background noise and speaker characteristics, among other information. Among these sources of variability, the current training of speech processing systems without regard to the impact of these sources will affect its performance and effectiveness. For example, only a portion of this information is related to ASR, while the rest can be considered as invariant and therefore possibly impinge upon the performance of ASR systems. This effect may lead to gender-biased or race-specific systems~\cite{NIPS2019_9662}. Koenecke et al. in~\cite{koenecke2020racial} examine the racial disparities of five state-of-the-art ASR systems developed by Amazon, Apple, Google, IBM, and Microsoft by transcribing structured interviews conducted with 42 white speakers and 73 black speakers. They found that there are disparities in the underlying acoustic models used by these ASR systems and they do not work equally well for all subgroups of the population. Likewise, the implementation of disentanglement in learning speaker representations can enhance the robustness of speaker representations and overcome common speaker recognition issues like anti-spoofing~\cite{peri2020empirical}.
Many recent applications have suggested that a disentangled speech representation can improve the interpretability and transferability of the representation in the speech signal~\cite{hsu2017unsupervised}. Although these applications seek to improve the quality and effectiveness of speech processing systems, it has not been considered for use in protecting privacy. We observe that the ability of the proposed framework to disentangle these representations can reconstruct different outputs that reflect a variety of privacy preferences. Thus, it can be argued that the separation of these representations will help to develop future privacy-aware solutions between users and service providers. Moreover, learning disentangled representations that reflect users’ preferences can bring enhanced robustness, interpretability, and controllability. We will, in future, seek to combine different techniques like adversarial training~\cite{huang2020unsupervised} and Siamese networks~\cite{last2020unsupervised} with disentanglement, or add further constraints grounded in information theory, to improve learning such disentangled representations from users' signals.

\textbf{Can we really do this at the edge?}
One of the primary reasons for taking an edge computing approach is to filter data locally prior to sending it to the cloud. Local filtering may be used to enhance protection of users' privacy. For example, an on-device transformation of sensor data was proposed by Malekzadeh et al. in~\cite{malekzadeh2019mobile}. They used convolutional autoencoders (CAE) as a sensor data anonymizer to remove user identifiable features locally and then share the filtered sensor data with specific applications, such as those designed for daily activities monitoring. In this work, we show how urgent it is to develop on-device privacy-preserving solutions for voice inputs by extracting the distinguishing representation from the speech without compromising individual privacy.
In earlier versions of this work~\cite{emotionless_2019}, we developed a privacy-preserving filter for voice inputs on edge devices to protect private paralinguistic information of a speaker. This filter enables users to protect their sensitive attributes (e.g. emotion) while benefiting from sharing their voice data with cloud-based voice analysis services. We implemented and evaluated the on-device filtering approach using a Raspberry Pi 4 as an example of an edge device, and our experimental results showed that similar performance in protecting sensitive information is attainable at the edge in comparison with cloud-based approaches. Although we showed that it is feasible for such models to be run on edge devices, further work is required to improve their efficiency, particularly with regard to model size and execution time. For example, model execution on a Raspberry Pi 4 takes twice as long ($\sim$40 seconds) as the cloud. In this work, our prototype implementation indicates the effectiveness of the proposed framework in reconstructing the speech signal. In addition, there is a decrease in the model size from about 126 MB to 95 MB. As future work, we aim to significantly reduce the execution time and memory usage of running the proposed framework on edge devices by further optimizing and quantizing the implementation of the model to make it suitable for use in real-time applications.
\section{\uppercase{Related Work}}
\label{sec:relatedwork}

\textbf{Privacy Leakage in Deep Learning.}
Deep learning models are vulnerable to various inference attacks as they remember information about their training data. Unwanted learning in the deep learning models was indicated by~\cite{song2017machine, melis2019exploiting}, showing that models leak detailed information about their training datasets. Likewise, in~\cite{carlini2019secret}, it is shown that generative text models trained on sensitive data can memorize training data and an attacker could extract unique and secret sequences like credit card numbers given these models. Song et al. define ``overlearning'' on deep learning models to be a model trained for a simple objective that can be re-purposed for a privacy-violating task in~\cite{song2019overlearning}. Motivated by these previous works, and given the scarcity of works targeting speech processing models specifically underlying deep acoustic models, in this paper we demonstrate the privacy leakage of input data from these models.

Other works have focused on protecting against membership inference attacks, which aim to determine whether a given data sample is used in the model's training~\cite{shokri2017membership}. Nasr et al. measure training data privacy leakage of deep learning algorithms by analyzing state-of-the-art pre-trained models from the CIFAR dataset in~\cite{nasr2018comprehensive}. They show that even well-generalized deep models are exposed to white-box membership inference attacks and leak a significant amount of information about their training data. Investigating membership inference attack is, however, beyond the scope of this paper but worthy of further investigation. We focus instead on the scenario whereby attackers can infer a significant amount of private information by observing the model input even if it is not in the training data.

Attribute inference attacks have been shown to compromise user privacy in various application domains including recommender systems~\cite{217523}, side-channel attacks~\cite{wei2018know}, location inference attacks~\cite{shokri2012protecting}, and property inference attacks~\cite{ganju2018property}. In these attacks, an attacker aims to infer the private attributes of the target user from his/her public data. Ateniese et al. show how an attacker can use access to the parameters of machine learning models such as Hidden Markov Models (HMM) to extract a predicate of the training data (e.g., the accent of the speaker in speech recognition models)~\cite{ateniese2015hacking}. In contrast to their work, we attest that such attacks perform well on the state-of-art underlying deep acoustic models for speech processing tasks to extract user-specific private attributes.

\textbf{Privacy Preserving Speech Representation.}
Learning privacy preserving representations in speech data is relatively unexplored~\cite{latif2020deep}. In~\cite{nautsch-2019} Nautsch et al. investigate the importance of the development of privacy-preserving technologies to protect speech signals and highlight the importance of applying these technologies to protect speakers and speech characterization in recordings.
Some recent works have sought to protect speaker identity~\cite{Hidebehind_2018}, gender identity~\cite{jaiswal2019privacy} and emotion~\cite{emotionless_2019}. VoiceMask, for example, was proposed to mitigate the security and privacy risks of voice input on mobile devices by concealing voiceprints~\cite{Hidebehind_2018}. It aims to strengthen users’ identity privacy by sanitizing the voice signal received from the microphone and then sending the perturbed speech to the voice input apps or the cloud. Moreover, in~\cite{emotionless_2019} an edge-based system is proposed to filter affect patterns from a user's voice before sharing it with cloud services for further analysis.
Unlike other approaches, however, we seek to protect the privacy of multiple user attributes for IoT scenarios that depend on voice input or speech analysis, i.e. sanitizing the speech signal of attributes a user may not wish to share but without decreasing functionality. We also emphasize the importance of learning disentangled speech representation for optimizing the privacy-utility trade-off and promoting privacy in a transparent manner.

\textbf{Fairness Representation.}
Fairness in machine learning is related to this work and shares similar methods, but where the objective is not to protect privacy. It aims to develop models that are invariant to particular attributes such as demographic information~\cite{madras2018learning}. In~\cite{edwards2015censoring} the authors have shown how the adversarial approach can be adapted to the task of removing sensitive information from representations. In~\cite{moyer2018invariant}, Moyer et al. have argued, however, that adversarial training for fairness and invariance is unnecessary, and sometimes produces counterproductive effects. Disentanglement has recently been shown to be useful for learning and evaluating fair machine learning models. Creager et al. proposed a fair representation learning model by disentanglement to achieve subgroup fairness in~\cite{creager2019flexibly}. Similarly, Locatello et al. investigated how disentanglement impacts the fairness of general-purpose representations in~\cite{locatello2019fairness}. In~\cite{NIPS2019_8699}, disentangling influence was presented to learn the influence of such attributes in accomplishing a given task. The authors investigate the importance of a feature's influence over the model outcomes taking advantage of disentangled representations. By contrast, our goal is to protect user privacy by preventing attackers from obtaining sensitive information, which is significantly different from the motivation and goals of previous studies.
\section{\uppercase{Conclusion}}
\label{sec:conclusion}

In this paper, we demonstrated vulnerabilities of underlying acoustic models used by speech processing tasks under attribute inference attacks. We proposed a privacy-aware, configurable framework for optimizing data sharing through voice user interfaces. Our proposed framework works in two phases, where the first phase adjusts privacy preferences and the second filters out sensitive attributes from users' input data depending on the configured privacy preference. We based our evaluation on empirical results derived from numerous real-world datasets, and show that the proposed framework can effectively defend against this class of attack. Specifically, we can reduce the success rate of inferring private attributes to less than or equal to chance, while providing on average over 99\% accuracy in primary tasks. In the next steps of the work, we intend to focus on extending our framework to be more \emph{tunable} to provide users with more controls depending on the devices and services with which they are interacting. An interesting direction for future research is to explore new privacy-preserving applications that can be enabled by the interpretability and controllability brought about by disentangled representations. 

\section*{\uppercase{ACKNOWLEDGEMENTS}}
This research was part-funded by the Saudi Arabian Cultural Bureau in the UK and EPSRC Databox and DADA grants (EP/N028260/1, EP/R03351X/1).

\bibliographystyle{ACM-Reference-Format}
\balance
\bibliography{RanyaRefs}

\end{document}